\documentclass[extra,referee]{gji}
\usepackage{timet}
\usepackage{amsmath, enumitem}
\usepackage{amsfonts, bm}
\usepackage{upgreek}
\usepackage{graphicx}
\graphicspath{{Figures/}} 
\usepackage{setspace}
\usepackage{multirow}
\usepackage[T1]{fontenc}
\usepackage{fixltx2e}
\usepackage{xcolor}
\usepackage[skins,breakable]{tcolorbox}
\usepackage{caption}

\newtcolorbox{cvbox}[2][]{%
  blanker,
  width = 0.9\textwidth,
  after skip=8mm,
  title=#2,
  breakable,
  #1
}

\DeclareMathOperator*{\argmax}{arg\,max}
\DeclareMathOperator*{\argmin}{arg\,min}

\begin{document}

\begin{titlepage}
    \begin{center}
        \vspace*{-2.0cm} 
        

		 \vspace{5cm}
		 
        \huge
        \textbf{3D Bayesian Variational Full Waveform Inversion}
        
        \vspace{4.0cm}
        \LARGE
        Xin Zhang$^1$, Angus Lomas$^{2}$, Muhong Zhou$^{2}$, \\
        York Zheng$^{2}$ and Andrew Curtis$^{1}$ \\

        \vspace{1cm}
        \Large
        $^1$ School of GeoSciences, University of Edinburgh, UK \\
        $^2$ BP p.l.c., London, UK \\
        
        \vspace{1cm}
        \Large
        E-mail: \textit{x.zhang2@ed.ac.uk, andrew.curtis@ed.ac.uk}

        \vfill
        \vfill
    \end{center}
    
\end{titlepage}

\newpage

%
%
%
%
\begin{summary}
Seismic full-waveform inversion (FWI) provides high resolution images of the subsurface by exploiting information in the recorded seismic waveforms. This is achieved by solving a highly nonnlinear and nonunique inverse problem. Bayesian inference is therefore used to quantify uncertainties in the solution. Variational inference is a method that provides probabilistic, Bayesian solutions efficiently using optimization. The method has been applied to 2D FWI problems to produce full Bayesian posterior distributions. However, due to higher dimensionality and more expensive computational cost, the performance of the method in 3D FWI problems remains unknown. We apply three variational inference methods to 3D FWI and analyse their performance. Specifically we apply automatic differential variational inference (ADVI), Stein variational gradient descent (SVGD) and stochastic SVGD (sSVGD), to a 3D FWI problem, and compare their results and computational cost. The results show that ADVI is the most computationally efficient method but systematically underestimates the uncertainty. The method can therefore be used to provide relatively rapid but approximate insights into the subsurface together with a lower bound estimate of the uncertainty. SVGD demands the highest computational cost, and still produces biased results. In contrast, by including a randomized term in the SVGD dynamics, sSVGD becomes a Markov chain Monte Carlo method and provides the most accurate results at intermediate computational cost. We thus conclude that 3D variational full-waveform inversion is practically applicable, at least in small problems, and can be used to image the Earth's interior and to provide reasonable uncertainty estimates on those images.
\end{summary}

\section{Introduction}
A wide variety of academic studies and practical applications require that we interrogate the Earth's subsurface for answers to scientific questions. A common approach is to image subsurface properties in three dimensions using data recorded on the Earth's surface, and to interpret those images to address questions of interest. In order to provide well justified and robust answers to such interrogation problems, it is necessary to assess the uncertainty in property estimates \citep{arnold2018interrogation}.

Seismic full-waveform inversion (FWI) uses full seismic recordings to characterize properties of the Earth's interior, and can provide high resolution images of the subsurface \citep{tarantola1984inversion, gauthier1986two, tarantola1988theoretical, pratt1999seismic, tromp2005seismic, fichtner2006adjoint, plessix2006review}. The method has been applied at industrial scale \citep{virieux2009overview, prieux2013multiparameter, warner2013anisotropic}, regional scale \citep{chen2007full, fichtner2009full, tape2009adjoint, chen2015multiparameter}, and global scale \citep{french2014whole, bozdaug2016global, fichtner2018collaborative, lei2020global}.

Due to the nonlinearity of relationships between model parameters and seismic waveforms, insufficient data coverage and noise in the data, FWI always has nonunique solutions and infinitely many sets of model parameters fit the data to within their uncertainty. It is therefore important to quantify uncertainties in the solution in order to better assess the reliability of inverted models \citep{tarantola2005inverse}.

FWI problems are traditionally solved using optimization methods in which one seeks an optimal set of parameter values by minimizing the difference or misfit between observed data and model-predicted data. The strong nonlinearity and nonuniqueness of the problem implies that a good starting model is required to avoid convergence to incorrect solutions (generally alternative modes or stationary points of the misfit function). Such models are not always available in practice. To alleviate this requirement a range of misfit functions that may reduce multimodality have been proposed \citep{luo1991wave, gee1992generalized, fichtner2008theoretical, brossier2010data, van2010correlation, bozdaug2011misfit, metivier2016measuring, warner2016adaptive, yuan2020exponentiated, sambridge2022geophysical}. Nevertheless, none of the standard methods of solution using any of these misfit functions have been shown to allow accurate estimates of uncertainty to be made in realistic FWI problems.

Bayesian inference provides a different way to solve inverse problems and quantify uncertainties. The method uses Bayes' theorem to update a \textit{prior} probability density function (pdf) with new information from the data to construct a so-called \textit{posterior} probability density function. The prior pdf describes information available about the parameters of interest prior to the inversion (independently of the current data set), while the posterior pdf describes the resultant state of information after combining information in the prior pdf with information in the current data. In principle, Bayesian inference thus provides accurate estimates of uncertainty.

Markov chain Monte Carlo (McMC) is one method to characterize the posterior pdf which has been used widely in many fields. In McMC one constructs a set (chain) of successive samples generated from the posterior pdf by taking a structured random walk in parameter space \cite[e.g.,][]{brooks2011handbook}; those samples can thereafter be used to infer the values of useful statistics of that pdf (mean, standard deviation, etc.). The Metropolis-Hastings algorithm is one such method \citep{metropolis1949monte, hastings1970monte} and has been applied to many applications in geophysics, including gravity inversion \citep{mosegaard1995monte, bosch2006joint, rossi2017bayesian}, vertical seismic profile inversion \citep{malinverno2000monte}, surface wave dispersion inversion \citep{bodin2012transdimensional, shen2012joint, young2013transdimensional, galetti2017transdimensional, zhang20183}, electrical resistivity inversion \citep{malinverno2002parsimonious, galetti2018transdimensional}, electromagnetic inversion \citep{minsley2011trans, ray2013robust, blatter2019bayesian}, travel time tomography \citep{bodin2009seismic, galetti2015uncertainty, galetti2017transdimensional} and more recently full-waveform inversion \citep{ray2017low, sen2017transdimensional, guo2020bayesian}. However, the basic Metropolis-Hastings algorithm becomes computationally intractable in high dimensional space  if the chain is attracted to individual misfit minima rather than exploring all possible such minima. To reduce this issue, more advanced McMC methods have been introduced to geophysics, such as Hamiltonian Monte Carlo \citep{duane1987hybrid, fichtner2018hamiltonian, gebraad2020bayesian, kotsi2020time}, stochastic Newton McMC \citep{martin2012stochastic, zhao2019gradient}, Langevin Monte Carlo \citep{roberts1996exponential, siahkoohi2020uncertainty} and parallel tempering \citep{hukushima1996exchange, dosso2012parallel, sambridge2013parallel}. However, the above studies mainly address 1D or 2D problems because of the high computational expense of moving to 3D. Some studies have applied McMC methods to 3D inverse problems including body wave travel time tomography \citep{piana2015local, hawkins2015geophysical, burdick2017velocity, zhang2020imaging} and surface wave dispersion inversion \citep{zhang20183, zhang20201d, ryberg2022ambient}, but they require enormous computational cost even for small datasets. Thus, McMC methods are generally considered to be intractable for large datasets and high dimensionality, such as occurs in 3D FWI problems.

Variational inference solves Bayesian inference problems in a different way: within a predefined family of (simplified) pdfs, the method seeks an optimal approximation to the posterior pdf by minimizing the difference between the approximating pdf and the posterior pdf. A typical metric used to measure this difference is the Kullback-Leibler (KL) divergence \citep{kullback1951information}. The method therefore solves an optimization problem rather than a stochastic sampling process as in McMC methods. As a result, in some classes of problems variational inference may be computationally more efficient than McMC methods and provide better scaling to higher dimensionality \citep{bishop2006pattern, blei2017variational, zhang2018advances}. The method can be applied to larger datasets by dividing the dataset into small minibatches and using stochastic and distributed optimization methods \citep{robbins1951stochastic, kubrusly1973stochastic}. In addition, the method can usually be parallelized at the individual sample level which makes the method even more efficient in real time by taking advantage of modern high performance computational facilities. By contrast, McMC methods cannot be parallelized at the sample level since each sample depends on the previous sample, and cannot use minibatches as these break the detailed balance condition that is required by common McMC methods \citep{o2004kendall}.

In variational inference the choice of variational family is important as it determines the accuracy of the approximation and the complexity of the optimization problem. A good choice should be rich enough to approximate complex distributions and simple enough such that the optimization problem remains solvable. Difference choices of variational families lead to a variety of specific methods. For example, a common choice is to use a \textit{mean-field} approximation in which the parameters are assumed to be mutually independent \citep{bishop2006pattern,blei2017variational}. In geophysics the method has been applied to invert for geological facies distributions using seismic data \citep{nawaz2018variational, nawaz2019rapid, nawaz2020variational}. While often leading to highly efficient algorithms, this method usually requires bespoke mathematical derivations which restricts its applicability to a limited range of problems. Based on a Gaussian variational family, \cite{kucukelbir2017automatic} proposed a method called automatic differential variational inference (ADVI), which can be applied easily to general problems. For example, the method has been used to solve seismic travel time tomography \citep{zhang2020seismic} and earthquake slip inversion problems \citep{zhang2022bayesian}. 

By exploiting the properties of probability transformations, another set of methods has been proposed in which one optimizes a series of invertible transforms which convert a simple initial distribution to an arbitrary distribution that can approximate the posterior distribution \citep{rezende2015variational, tran2015variational, liu2016stein}. Normalizing flow variational inference is one such method which applies a series of invertible and differential transforms (called flows) to an initial distribution; those flows are then optimized to produce an improved approximation to the posterior pdf \citep{rezende2015variational}. Normalizing flows have been demonstrated to be an efficient method in geophysical applications such as seismic tomography \citep{zhao2021bayesian} and image denoising \citep{siahkoohi2020faster}. However, the method becomes inefficient in very high dimensional space because of the computational cost required by large and flexible forms of flows. Stein variational gradient descent (SVGD) provides an alternative method that uses a set of particles (models) to represent the probability distribution. Those particles are iteratively updated by minimizing the KL-divergence so that in their final state their density approximates the posterior pdf \citep{liu2016stein}. The method has been applied to a range of geophysical applications, including seismic travel time tomography \citep{zhang2020seismic}, earthquake location \citep{smith2022hyposvi}, hydrogeological inversion \citep{ramgraber2021non} and 2D full-waveform inversion \citep{zhang2020variational, zhang2021bayesiana}. However, none of these studies are comparable to a typical 3D FWI problem in terms of dimensionality and computational cost, so the property of the method in 3D FWI remains unknown.

In this study we explore the properties and efficiency of variational inference methods in 3D FWI problems, including ADVI and SVGD. In addition, to reduce possible deficiency of SVGD in higher dimensionality \citep{ba2021understanding} we introduce another method called stochastic SVGD \cite[sSVGD:][]{gallego2018stochastic} and compare the method with ADVI and SVGD. In section 2 we first describe the basic concept of variational inference and then the ADVI, SVGD and sSVGD methods. In section 3 we apply the suite of methods to a 3D FWI problem and compare their results and computational costs. The aim of this study is to explore performance of those methods, to assess the computational requirements and to provide useful information for practitioners. Our results demonstrate that the 3D variational FWI is practically feasible, at least for small problems, and so can be applied to image the Earth's subsurface and to provide uncertainty estimates on the results.

\section{Methods}
\subsection{Variational inference}
Bayesian inference is the process of constructing a posterior probability density function $p(\mathbf{m}|\mathbf{d}_{\mathrm{obs}})$ of model parameters $\mathbf{m}$ given the observed data $\mathbf{d}_{\mathrm{obs}}$, by updating a prior pdf with new information contained in the data. According to Bayes' theorem,
\begin{equation}
    p(\mathbf{m}|\mathbf{d}_{\mathrm{obs}}) = \frac{p(\mathbf{d}_{\mathrm{obs}}|\mathbf{m})p(\mathbf{m})}{p(\mathbf{d}_{\mathrm{obs}})}
\label{eq:Bayes}
\end{equation}
where $p(\mathbf{d}_{\mathrm{obs}}|\mathbf{m})$ is the \textit{likelihood} which describes the probability of observing data $\mathbf{d}_{\mathrm{obs}}$ if model $\mathbf{m}$ was true, $p(\mathbf{m})$ represents the prior pdf which describes information that is known independently of the data, and $p(\mathbf{d}_{\mathrm{obs}})$ is a normalization factor called the \textit{evidence}. 

Variational inference solves the above Bayesian inference problem using optimization. The method seeks an optimal approximation $q^{*}(\mathbf{m})$ to the posterior pdf $p(\mathbf{m}|\mathbf{d}_{\mathrm{obs}})$ within a predefined family of known probability distributions $Q=\{q(\mathbf{m})\}$ by minimizing the KL divergence between $q(\mathbf{m})$ and $p(\mathbf{m}|\mathbf{d}_{\mathrm{obs}})$:
\begin{equation}
	q^{*}(\mathbf{m}) = \argmin_{q \in Q} \mathrm{KL}[q(\mathbf{m})||p(\mathbf{m}|\mathbf{d}_{\mathrm{obs}})]
	\label{eq:argmin_KL} 
\end{equation}
The KL divergence measures the difference between two probability distributions and can be expressed as:
\begin{equation}
\mathrm{KL}[q(\mathbf{m})||p(\mathbf{m}|\mathbf{d}_{\mathrm{obs}})] = \mathrm{E}_{q}[\mathrm{log}q(\mathbf{m})] - \mathrm{E}_{q}[\mathrm{log}p(\mathbf{m}|\mathbf{d}_{\mathrm{obs}})]
\label{eq:KL}
\end{equation}
where the expectation is taken with respect to the distribution $q(\mathbf{m})$. The KL divergence is nonnegative and only equals zero when $q(\mathbf{m}) = p(\mathbf{m}|\mathbf{d}_{\mathrm{obs}})$ \citep{kullback1951information}. Expanding the posterior pdf using equation (\ref{eq:Bayes}), the KL divergence becomes:
\begin{equation}
\mathrm{KL}[q(\mathbf{m})||p(\mathbf{m}|\mathbf{d}_{\mathrm{obs}})] = \mathrm{E}_{q}[\mathrm{log}q(\mathbf{m})] - \mathrm{E}_{q}[\mathrm{log}p(\mathbf{m},\mathbf{d}_{\mathrm{obs}})] + \mathrm{log}p(\mathbf{d}_{\mathrm{obs}})
\label{eq:KL2}
\end{equation}
The evidence term $\mathrm{log}p(\mathbf{d}_{\mathrm{obs}})$ is computationally intractable because it requires evaluation of a high dimensional integral for which the computation scales exponentially with the number of parameters. We therefore rearrange equation (\ref{eq:KL2}) to obtain the evidence lower bound (ELBO):
\begin{equation}
\begin{aligned}
\mathrm{ELBO}[q]  & = \mathrm{log}p(\mathbf{d}_{\mathrm{obs}}) - \mathrm{KL}[q(\mathbf{m})||p(\mathbf{m}|\mathbf{d}_{\mathrm{obs}})] \\
& =  \mathrm{E}_{q}[\mathrm{log}p(\mathbf{m},\mathbf{d}_{\mathrm{obs}})] - \mathrm{E}_{q}[\mathrm{log}q(\mathbf{m})]
\end{aligned}
\label{eq:ELBO}
\end{equation}
Since the KL divergence is nonnegative, the above equation defines a lower bound for the evidence $\mathrm{log}p(\mathbf{d}_{\mathrm{obs}})$. In addition because the evidence $\mathrm{log}p(\mathbf{d}_{\mathrm{obs}})$ is a constant for a given problem, minimizing the KL-divergence is equivalent to maximizing the ELBO. Consequently, variational inference in equation (\ref{eq:argmin_KL}) can also be expressed as:
\begin{equation}
	q^{*}(\mathbf{m}) = \argmax_{q \in Q} \mathrm{ELBO}[q(\mathbf{m})]
	\label{eq:argmax_elbo} 
\end{equation}

In variational inference, the choice of the variational family $Q$ is important because it determines the accuracy of the approximation as well as the complexity of the optimization problem. Different methods can be developed depending on difference choices of the family. In the following sections we describe a set of different methods: ADVI, SVGD and sSVGD, and compare these methods in the application of 3D full-waveform inversion.
 
\subsection{Automatic differential variational inference (ADVI)}

\begin{figure}
	\includegraphics[width=1.\linewidth]{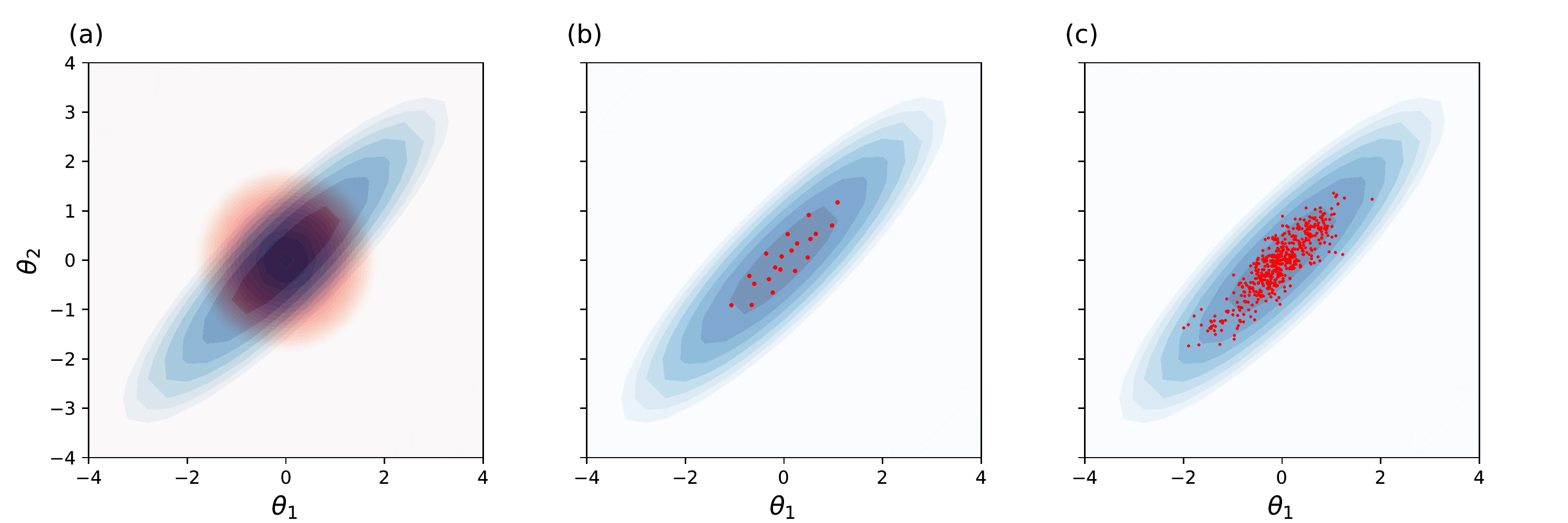}
	\caption{(a) The posterior distribution (red) obtained using ADVI with a mean-field approximation, and the samples obtained using (b) SVGD and (c) sSVGD in the case of a bivariate Gaussian distribution (blue). For both SVGD and sSVGD 20 particles are used.}
	\label{fig:mean-field}
\end{figure}

ADVI is a variational method based on a Gaussian variational family \citep{kucukelbir2017automatic}. Gaussians are defined on the entire set of real numbers and in reality model parameters often have hard constrains (for example, seismic velocity is greater than zero), so in ADVI we first transform those constrained parameters into an unconstrained space using an invertible transform $T: \bm{\uptheta} = T(\mathbf{m})$. In this space the joint probability $p(\mathbf{m},\mathbf{d}_{\mathrm{obs}})$ becomes: 
\begin{equation}
 p(\bm{\uptheta},\mathbf{d}_{\mathrm{obs}}) = p(\mathbf{m},\mathbf{d}_{\mathrm{obs}})|det\mathbf{J}_{T^{-1}}(\bm{\uptheta})|
 \label{eq:prob_transforma}
\end{equation}
where $\mathbf{J}_{T^{-1}}(\bm{\uptheta})$ is the Jacobian matrix of the inverse of $T$ and $|\cdot|$ denotes absolute value. Define a Gaussian variational family
\begin{equation}
q(\bm{\uptheta};\bm{\zeta})=\mathcal{N}(\bm{\uptheta}|\bm{\upmu},\bm{\Sigma})
\label{eq:normal}
\end{equation}
where $\bm{\zeta}$ represents variational parameters, that is the mean vector $\bm{\upmu}$ and the covariance matrix $\bm{\Sigma}$. Although a full covariance matrix can be used for small size problems, it becomes computationally intractable for very high dimensional space (as in 3D FWI). We therefore use a factorized (mean-field) Gaussian variational approximation:
\begin{equation}
	q(\bm{\uptheta};\bm{\zeta})=\mathcal{N}(\bm{\uptheta}|
	\bm{\upmu},\mathrm{diag} ({\mathrm{exp}(\bm{\omega})^{2}}) )
	\label{eq:meanfield_normal}
\end{equation}
where we have reparameterized the standard deviation using $\bm{\sigma}=\mathrm{exp}(\bm{\omega})$ to ensure that each parameter of $\bm{\sigma}$ is positive. Note that because we neglect the correlation information between different parameters, the approximation obtained by minimizing the KL divergence systematically underestimates the marginal variance as illustrated in Figure \ref{fig:mean-field}a \citep{bishop2006pattern}. 

With the above definition the variational problem in equation (\ref{eq:argmax_elbo}) can be written as:
\begin{equation}
\begin{aligned}
 \bm{\zeta}^{*} &= \argmax_{\bm{\zeta}} \mathrm{ELBO}[q(\bm{\uptheta};\bm{\zeta})]\\
 &= \argmax_{\bm{\zeta}} \mathrm{E}_{q}\left[\mathrm{log} p( T^{-1}(\bm{\uptheta}),\mathbf{d}_{\mathrm{obs}} )+\mathrm{log}|det\mathbf{J}_{T^{-1}}(\bm{\uptheta})| \right] - \mathrm{E}_{q} \left[ \mathrm{log}q(\bm{\uptheta}) \right] 
 \end{aligned}
 \label{eq:argmaxELBO}
\end{equation}
This optimization problem can be solved by using gradient ascent methods. As shown in \cite{kucukelbir2017automatic} the gradients of the ELBO with respect to $\bm{\upmu}$ and $\bm{\omega}$ are:
\begin{equation}
 \nabla_{\bm{\upmu}}\mathrm{ELBO} = \mathrm{E}_{\mathcal{N}(\bm{\eta}|\bm{0},\mathbf{I})}\left[\nabla_{\mathbf{m}}\mathrm{log}p(\mathbf{m},\mathbf{d}_{\mathrm{obs}})\nabla_{\bm{\theta}}T^{-1}(\bm{\uptheta})+\nabla_{\bm{\uptheta}}\mathrm{log}|det\mathbf{J}_{T^{-1}}(\bm{\uptheta})| \right]
 \label{eq:gradient_mu}
\end{equation}
\begin{equation}
 \nabla_{\bm{\omega}}\mathrm{ELBO} = \mathrm{E}_{\mathcal{N}(\bm{\eta}|\bm{0},\mathbf{I})}\left[ \left( \nabla_{\mathbf{m}}\mathrm{log} p(\mathbf{m},\mathbf{d}_{\mathrm{obs}})\nabla_{\bm{\uptheta}}T^{-1}(\bm{\uptheta})+\nabla_{\bm{\uptheta}}\mathrm{log}|det\mathbf{J}_{T^{-1}}(\bm{\uptheta})| \right) \bm{\eta}^{\mathrm{T}}\mathrm{diag}(\mathrm{exp}(\bm{\omega})) \right] + \bm{1}
 \label{eq:gradient_sigma}
\end{equation}
where $\bm{\eta}$ is a random variable generated from a standard Normal distribution $\mathcal{N}(\bm{0},\mathbf{I})$. The expectations in the above equations can be estimated by Monte Carlo (MC) integration, which in practice only requires a low number of samples because the optimization is usually performed over many iterations so that statistically the gradients will lead to convergence toward the correct solution \citep{kucukelbir2017automatic}. The variational problem in equation (\ref{eq:argmaxELBO}) can therefore be solved by using gradient ascent methods. The final approximation $q^{*}(\mathbf{m})$ is obtained by transforming $q^{*}(\bm{\uptheta})$ back to the original space. For the transform $T$, we use a commonly-employed logarithmic transform \citep{stan2016stan, zhang2020seismic}
\begin{equation}
\begin{aligned}
 \theta_{i} &= T(m_{i}) = \mathrm{log}(m_{i}-a_{i}) - \mathrm{log}(b_{i}-m_{i}) \\
 m_{i} &= T^{-1}(\theta_{i}) = a_{i} + \frac{(b_{i}-a_{i})}{1+exp(-\theta_{i})}
\end{aligned}
 \label{eq:transform}
\end{equation} 
where $m_{i}$ represents $i^{th}$ parameter in the original constrained space, $\theta_{i}$ is the transformed variable in the unconstrained space, $a_{i}$ and $b_{i}$ are the lower and upper bound on $m_{i}$, respectively. Although ADVI can generate biased results as we discussed above, it has been demonstrated to be a computationally efficient method compared to SVGD \citep{zhang2020seismic, zhao2021bayesian}. For this reason, we explore its properties in 3D FWI problems.

\subsection{Stein variational gradient descent (SVGD)}

SVGD is a variational method which uses a set of samples (called particles) whose density represents the approximation pdf $q$. The method iteratively updates those particles by minimizing the KL divergence so that the final set of particles are distributed according to the posterior distribution \citep{liu2016stein}. Since the distribution of a set of particles is in principle entirely flexible, this method can provide more accurate results than ADVI \citep{zhang2020seismic}. Define the set of particles as $\{\mathbf{m}_{i}\}$ where $\mathbf{m}_{i}$ is a $d$-dimensional parameter vector. SVGD uses a smooth transform $T(\mathbf{m}_{i})=\mathbf{m}_{i} + \epsilon\bm{\upphi}(\mathbf{m}_{i})$ to update each particle, where $\bm{\upphi} = [\upphi_{1},...,\upphi_{d}]$ is a smooth vector function that describes the perturbation direction and $\epsilon$ is the magnitude of the perturbation. Assume $T$ is invertible and define $q_{T}(\mathbf{m})$ as the transformed probability distribution of pdf $q(\mathbf{m})$. The gradient of the KL-divergence between $q_{T}$ and the posterior pdf $p$ with respect to $\epsilon$ can be computed as \citep{liu2016stein}:
\begin{equation}
\nabla_{\epsilon} \mathrm{KL}[q_{T}||p] \, |_{\epsilon=0} = - \mathrm{E}_{q} 
\left[ trace \left( \mathcal{A}_{p} \bm{\upphi} (\mathbf{m}) \right) \right] 
\label{eq:stein_gradient}
\end{equation} 
where $\mathcal{A}_{p}$ is the Stein operator defined by $\mathcal{A}_{p} \bm{\upphi}(\mathbf{m}) = \nabla_{\mathbf{m}} \mathrm{log} p(\mathbf{m}) \bm{\upphi} (\mathbf{m})^{T} + \nabla_{ \mathbf{m} } \bm{\upphi} ( \mathbf{m} )$. Equation \ref{eq:stein_gradient} ensures that by maximizing the right-hand expectation we obtain the steepest descent direction of the KL-divergence, and consequently the KL divergence can be minimized by iteratively stepping a small distance in that direction.

The optimal $\bm{\upphi}^{*}$ that maximize the expectation in equation (\ref{eq:stein_gradient}) can be found by using kernel functions. Say $x,y \in X$ and define a mapping $\psi$ from $X$ to a space where an inner product $\langle,\rangle$ is defined (called a Hilbert space); a \textit{kernel} is a function that satisfies $k(x,y)=\langle \psi(x),\psi(y) \rangle$. Assume a kernel $k(\mathbf{m}',\mathbf{m})$, the optimal $\bm{\upphi}^{*}$ can be expressed as \citep{liu2016stein}:
\begin{equation}
	\bm{\upphi}^{*} \propto \mathrm{E}_{\{\mathbf{m'} \sim q\}} [\mathcal{A}_{p} k(\mathbf{m'},\mathbf{m})]
	\label{eq:phi_star}
\end{equation}
Since we use particles $\{\mathbf{m}_{i}\}$ to represent $q$, the expectation can be approximated using the particles mean. The KL divergence can therefore be minimized by iteratively applying the transform $T(\mathbf{m})=\mathbf{m}+\epsilon\bm{\upphi}^{*}(\mathbf{m})$ to a set of initial particles $\{\mathbf{m}_{i}^{0}\}$: 
\begin{equation}
	\begin{aligned}
		\bm{\upphi}^{*}_{l} (\mathbf{m}) &= \frac{1}{n} \sum_{j=1}^{n} \left[ k(\mathbf{m}_{j}^{l} , \mathbf{m}) \nabla_{\mathbf{m}_{j}^{l}} \mathrm{log} p(\mathbf{m}_{j}^{l}|\mathbf{d}_{\mathrm{obs}}) + \nabla_{\mathbf{m}_{j}^{l}} k(\mathbf{m}_{j}^{l}, \mathbf{m}) \right] \\
		\mathbf{m}_{i}^{l+1} &= \mathbf{m}_{i}^{l} + \epsilon^{l} \bm{\upphi}^{*}_{l} (\mathbf{m}_{i}^{l})
	\end{aligned}
	\label{eq:phi_mean}
\end{equation}
where $l$ denotes the $l^{th}$ iterations, $n$ is the number of particles and $\epsilon^{l}$ is the step size. If the step size $\{\epsilon^{l}\}$ is sufficiently small then the transform $T$ is invertible, and the process converges to the posterior pdf asymptotically as the number of particles tends to infinity.

For the kernel function we use a commonly-used radial basis function (RBF)
\begin{equation}
	k(\mathbf{m},\mathbf{m}') = \mathrm{exp} [- \frac{\Vert \mathbf{m}-\mathbf{m}' \Vert^{2}}{2h^{2}}]
	\label{eq:rbf}
\end{equation}
where $h$ is a scale factor which intuitively controls the interaction intensity between different particles based on their distances apart. As suggested by several studies \citep{liu2016stein, zhang2020seismic, zhang2020variational}, we choose $h$ to be $ \tilde{d} / \sqrt{2\mathrm{log}n}$ where $\tilde{d}$ is the median of pairwise distances between all particles. This choice ensures that the contribution from each particle $\mathbf{m}_{i}$'s own gradient is balanced by the influence from all other particles as $\sum_{j \ne i}k(\mathbf{m}_{i}, \mathbf{m}_{j}) \approx n\mathrm{exp}(-\frac{1}{2h^{2}} \tilde{d}^{2}) = 1$. Note that for the RBF kernel, the second term of $\bm{\upphi}^{*}$ in equation (\ref{eq:phi_mean}) becomes  $\sum_{j} \frac{\mathbf{m}-\mathbf{m}_{j}}{\sigma^{2}}  k(\mathbf{m}_{j}, \mathbf{m})$ which drives the particle $\mathbf{m}$ away from its neighbouring particles when the kernel takes high values. This second term therefore acts as a repulsive force which prevents the particles from collapsing to a single mode, whereas the first term consists of kernel weighted gradients which drives the particles toward high probability areas. An example of the particles obtained using SVGD in the case of a bivariate Gaussian distribution is shown in Figure \ref{fig:mean-field}b.

In Geophysics, SVGD has been demonstrated to be an efficient method for a rang of applications \citep{zhang2020seismic, zhang2020variational, zhang2021bayesiana, ramgraber2021non, zhao2021bayesian, smith2022hyposvi, ahmed2022regularized}. In this study we explore its applicability in 3D full-waveform inversion. As in previous studies \citep{zhang2020variational, zhang2021introduction}, in order to handle hard constraints of seismic velocity, we transform seismic velocity into an unconstrained space using equation (\ref{eq:transform}) and perform SVGD in that space. The final seismic velocities are obtained by transforming the particles back to the original space.

\subsection{Stochastic SVGD}
Although SVGD has been applied to many different applications \citep{gong2019quantile, zhang2020seismic, zhang2020variational, pinder2020stein}, the method can provide biased results and is known to underestimate variance for high dimensional problems because of the finite number of particles and the practical limitation of computational cost \citep{ba2021understanding}. In order to further improve accuracy of the method, efforts have been made to bridge the gap between variational inference and McMC methods. sSVGD is one such algorithm which turns SVGD into a Markov chain by adding a Gaussian noise term to the dynamics \citep{gallego2018stochastic}. By doing this one can start collecting many samples that represent the posterior pdf after a burn-in period instead of having to use a large number of particle from the beginning. In addition, the method guarantees asymptotic convergence to the posterior pdf as the number of iterations tends to infinity, which standard SVGD with a finite number of particles cannot achieve.

To introduce the sSVGD algorithm, we start from a stochastic differential equation (SDE):
\begin{equation}
	d\mathbf{z} = \mathbf{f}(\mathbf{z})dt + \sqrt{2\mathbf{D}(\mathbf{z})}d\mathbf{W}(t)
	\label{eq:sde}
\end{equation}
where $\mathbf{f}(\mathbf{z})$ is called the \textit{drift}, $\mathbf{W}(t)$ is a Wiener process, and $\mathbf{D}(\mathbf{z})$ is a positive semidefinite diffusion matrix. Generally all continuous Markov processes can be expressed as a SDE of the above form. If we denote the posterior distribution as $p(\mathbf{z})$, \cite{ma2015complete} proposed a SDE that converges to the distribution $p(\mathbf{z})$
\begin{equation}
	\mathbf{f}(\mathbf{z}) = \left[\mathbf{D}(\mathbf{z}) + \mathbf{Q}(\mathbf{z}) \right]\nabla \mathrm{log}p(\mathbf{z}) + \Gamma(\mathbf{z})
	\label{eq:drift}
\end{equation}
where $\mathbf{Q}(\mathbf{z})$ is a skew-symmetric curl matrix, and $\Gamma_{i}(\mathbf{z}) = \sum_{j=1}^{d} \frac{\partial}{\partial\mathbf{z}_{j}}(\mathbf{D}_{ij}(\mathbf{z}) + \mathbf{Q}_{ij}(\mathbf{z}))$ is a correction term which amends the bias.

If we discretize equation (\ref{eq:sde}) with equation (\ref{eq:drift}) using the Euler-Maruyama discretization, we obtain a practical algorithm:
\begin{equation}
	\mathbf{z}_{t+1} = \mathbf{z}_{t} + \epsilon_{t} \left[ \left(\mathbf{D}\left(\mathbf{z}_{t}\right) + \mathbf{Q}(\mathbf{z}_{t})\right)\nabla \mathrm{log}p(\mathbf{z}_{t}) + \Gamma(\mathbf{z}_{t}) \right] + \mathcal{N}(\mathbf{0},2\epsilon_{t}\mathbf{D}(\mathbf{z}_{t}))
	\label{eq:discretized_sde}
\end{equation}
where $\mathcal{N}(\mathbf{0},2\epsilon_{t}\mathbf{D}(\mathbf{z}_{t}))$ represents a Gaussian distribution. The gradient $\nabla \mathrm{log}p(\mathbf{z}_{t})$ can be computed using full data, or Uniformly randomly selected minibatch data subsets which results in a stochastic gradient. In either case the above process converges to the posterior distribution asymptotically as $\epsilon_{t} \rightarrow 0$ and $t \rightarrow \infty$ \citep{ma2015complete}. Matrix $\mathbf{D}(\mathbf{z})$ and $\mathbf{Q}(\mathbf{z})$ can be adjusted to obtain faster convergence to the posterior distribution. For example, by setting $\mathbf{D}=\mathbf{I}$ and $\mathbf{Q}=\mathbf{0}$ one obtains the stochastic gradient Langevin dynamics algorithm \citep{welling2011bayesian}. If we augment the state space $\mathbf{z}$ with a moment term $\mathbf{x}$ to obtain an augmented space $\overline{\mathbf{z}}=(\mathbf{z},\mathbf{x})$, and set $\mathbf{D}=\mathbf{0}$ and $\mathbf{Q}=\left( \begin{array}{cc} 
	\mathbf{0} & \mathbf{-I} \\ \mathbf{I} & \mathbf{0}
\end{array}\right)$, the stochastic Hamiltonian Monte Carlo (HMC) method can be derived \citep{chen2014stochastic}. 

For the set of particles $\{\mathbf{m}_{i}\}$ defined in the above section we can construct an augmented space $\mathbf{z}=(\mathbf{m}_{1}, \mathbf{m}_{2}, ..., \mathbf{m}_{n}) \in \mathbb{R}^{nd}$ by concatenating $n$ particles, and use equation (\ref{eq:discretized_sde}) to obtain a valid sampler that runs multiple ($n$) interacted chains:
\begin{equation}
		\mathbf{z}_{t+1} = \mathbf{z}_{t} + \epsilon_{t} [(\mathbf{D}(\mathbf{z}_{t}) + \mathbf{Q}(\mathbf{z}_{t}))\nabla \mathrm{log}p(\mathbf{z}_{t}) + \Gamma(\mathbf{z}_{t})] + \mathcal{N}(\mathbf{0},2\epsilon_{t}\mathbf{D}(\mathbf{z}_{t}))
	\label{eq:general_sgmc}
\end{equation}
where $\mathbf{D}, \mathbf{Q} \in \mathbb{R}^{nd\times nd}$ and $\nabla \mathrm{log}p, \Gamma \in \mathbb{R}^{nd}$. Define a matrix $\mathbf{K}$
\begin{equation}
	\mathbf{K} = \frac{1}{n} \begin{bmatrix}
		k(\mathbf{m}_{1},\mathbf{m}_{1})\mathbf{I}_{d\times d} & \dots & k(\mathbf{m}_{1},\mathbf{m}_{n})\mathbf{I}_{d\times d} \\
		\vdots & \ddots & \vdots \\
		k(\mathbf{m}_{n},\mathbf{m}_{1})\mathbf{I}_{d\times d} & \dots & k(\mathbf{m}_{n},\mathbf{m}_{n})\mathbf{I}_{d\times d} 
	\end{bmatrix}
\label{eq:matrixK}
\end{equation}
where $k(\mathbf{m}_{i},\mathbf{m}_{j})$ is a kernel function and $\mathbf{I}_{d \times d}$ is an identity matrix. According to the definition of kernel functions, the matrix $\mathbf{K}$ is positive definite \citep{gallego2018stochastic}. The standard SVGD algorithm in equation (\ref{eq:phi_mean}) can now be expressed in matrix form as
\begin{equation}
	\mathbf{z}_{t+1} = \mathbf{z}_{t} + \epsilon_{t} [\mathbf{K} \nabla \mathrm{log}p(\mathbf{z}_{t}) + \nabla \cdot \mathbf{K}]
	\label{eq:matrix_svgd}
\end{equation}  
which shows that SVGD can be regarded as a special case of equation (\ref{eq:general_sgmc}) with $\mathbf{D}_{\mathbf{K}}=\mathbf{K}$, $\mathbf{Q}_{\mathbf{K}}=\mathbf{0}$ and no noise term. By including the noise term, we construct a stochastic gradient McMC method with SVGD gradients, which we call stochastic SVGD:
\begin{equation}
	\mathbf{z}_{t+1} = \mathbf{z}_{t} + \epsilon_{t} [\mathbf{K} \nabla \mathrm{log}p(\mathbf{z}_{t}) + \nabla \cdot \mathbf{K}]
	+ \mathcal{N}(\mathbf{0},2\epsilon_{t}\mathbf{K})
	\label{eq:stochastic_svgd}
\end{equation} 
According to the discussion above, this process converges to the posterior distribution $p(\mathbf{z})= \prod_{i=1}^{n} p(\mathbf{m}_{i}|\mathbf{d}_{\mathrm{obs}})$ asymptotically. Note that when the number of particles is large enough, the noise term would be tiny according to equation (\ref{eq:matrixK}). Consequently in such case the method produces the same results as standard SVGD. 

In order to use equation (\ref{eq:stochastic_svgd}) to sample the posterior distribution, we need to draw samples from the Gaussian distribution $\mathcal{N}(\mathbf{0},2\epsilon_{t}\mathbf{K})$. This requires computing the lower triangular Cholesky decomposition of the $nd \times nd$ matrix $\mathbf{K}$, which can be computationally expensive. To compute the noise term efficiently, we define a block-diagonal matrix $\mathbf{D}_{\mathbf{K}}$
\begin{equation}
	\mathbf{D}_{\mathbf{K}} = \frac{1}{n} 
	\left[ \renewcommand\arraystretch{0.3}
	\begin{array}{ccc}
		\overline{\mathbf{K}} & & \\
		& \ddots & \\
		& & \overline{\mathbf{K}}
	\end{array}
	\right]
\label{eq:matrixDK}
\end{equation}
where $\overline{\mathbf{K}}$ is a $n \times n$ matrix with $\overline{\mathbf{K}}_{ij} = k(\mathbf{m}_{i},\mathbf{m}_{j})$. Notice that with this definition, $\mathbf{D}_{\mathbf{K}}$ can be constructed from $\mathbf{K}$ using $\mathbf{D}_{\mathbf{K}} = \mathbf{P}\mathbf{K}\mathbf{P}^{\mathrm{T}}$ where $\mathbf{P}$ is a permutation matrix
\begin{equation}
	\mathbf{P} = 
	\left[\arraycolsep=2pt \def\arraystretch{0.2}
	\begin{array}{ccc|ccc|c|ccc}
		1 & & & & & & & & & \\
		& & & 1 & & & & & &  \\
		& & & & & & \ddots & & & \\
		& & & & & & & 1 & & \\
		\hline
		& 1 & & & & & & & & \\
		& & & & 1 & & & & &  \\
		& & & & & & \ddots & & & \\
		& & & & & & & & 1 & \\
		\hline
		& \ddots & & & \ddots & & \ddots & & \ddots & \\
		\hline
		& & 1 & & & & & & & \\
		& & & & & 1 & & & &  \\
		& & & & & & \ddots & & & \\
		& & & & & & & & & 1 \\
	\end{array}
    \right]
\end{equation}
The action of this permutation matrix on a vector $\mathbf{z}$ rearranges the order of the vector from the basis where the particles are listed sequentially to that where the first coordinates of all particles are listed, then the second, etc. The noise term $\bm{\eta}$ can therefore be generated using 
\begin{equation}
	\begin{aligned}
	\bm{\eta} &\sim \mathcal{N}(\mathbf{0},2\epsilon_{t}\mathbf{K}) \\
			  &\sim \sqrt{2\epsilon_{t}} \mathbf{P}^{\mathrm{T}}\mathbf{P}\mathcal{N}(\mathbf{0},\mathbf{K}) \\
			  &\sim \sqrt{2\epsilon_{t}} \mathbf{P}^{\mathrm{T}} \mathcal{N}(\mathbf{0},\mathbf{D}_{\mathbf{K}}) \\
			  &\sim \sqrt{2\epsilon_{t}} \mathbf{P}^{\mathrm{T}} \mathbf{L}_{\mathbf{D}_\mathbf{K}}  \mathcal{N}(\mathbf{0},\mathbf{I})
	\end{aligned}
\label{eq:noise_term}
\end{equation}
where $\mathbf{L}_{\mathbf{D}_\mathbf{K}}$ is the lower triangular Cholesky decomposition of matrix $\mathbf{D}_{\mathbf{K}}$. Given that $\mathbf{D}_{\mathbf{K}}$ is a block-diagonal matrix, decomposition $\mathbf{L}_{\mathbf{D}_\mathbf{K}}$ can be calculated easily as we only need to calculate the lower triangular Cholesky decomposition of matrix $\overline{\mathbf{K}}$. Since in practice the number of particles $n$ is usually modest, evaluating the noise term is computationally negligible. We can now use equation (\ref{eq:stochastic_svgd}) to generate samples from the posterior distribution. An example of the samples obtained using sSVGD in the case of a bivariate Gaussian distribution is shown in Figure \ref{fig:mean-field}c.
\section{Results}
\begin{figure}
	\includegraphics[width=1.\linewidth]{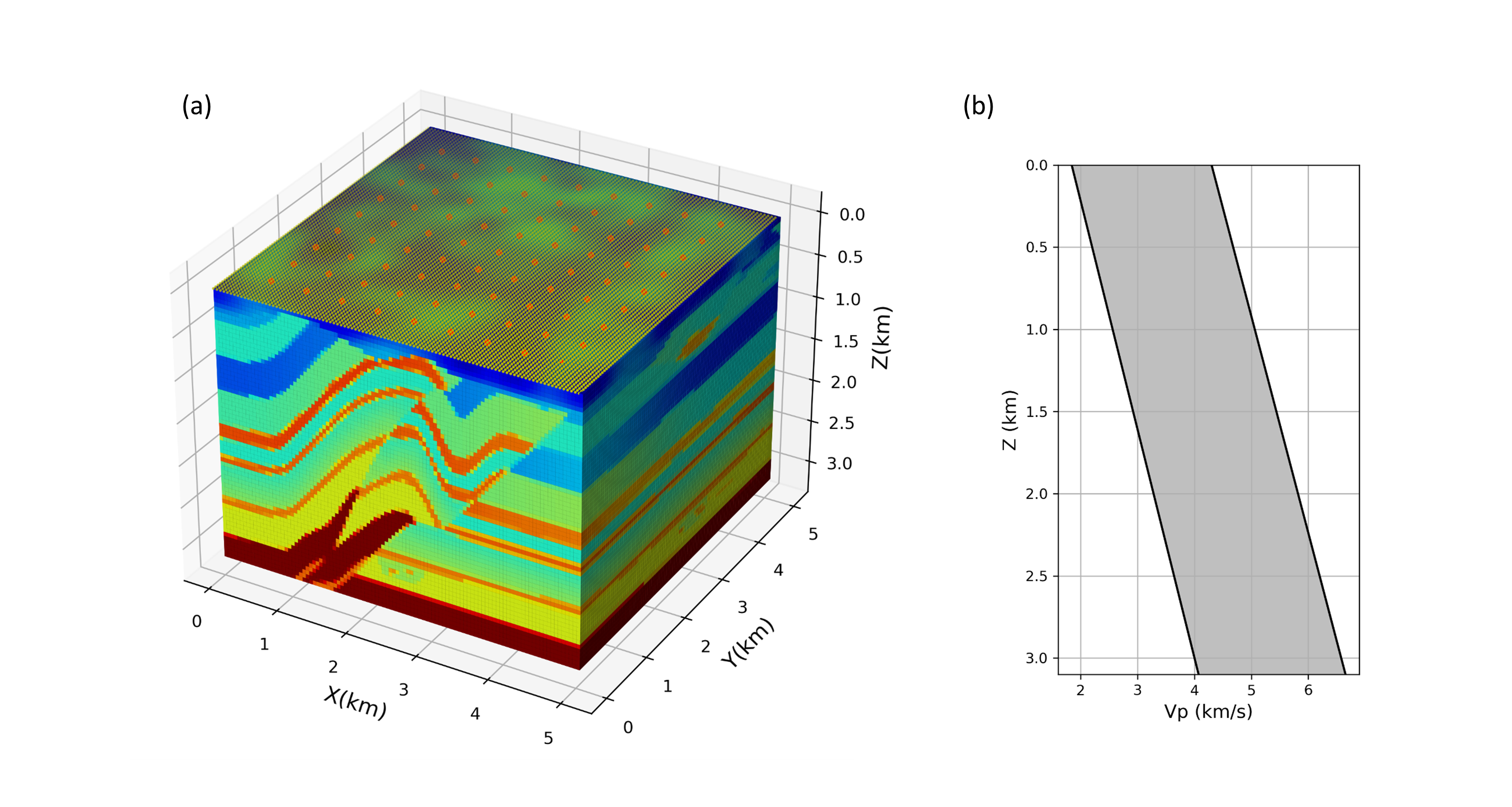}
	\caption{(\textbf{a}) True velocity model and acquisition geometry used in this study. Surface sources and receivers are denoted using red and yellow dots respectively. (\textbf{b}) Prior distribution used in the inversion: a Uniform distribution with a width of 2.5 km/s at each depth.}
	\label{fig:true_prior}
\end{figure}

We apply the above suite of methods to an acoustic 3D full-waveform inversion problem. The true model is chosen to be a part of the 3D overthrust model \cite[Figure \ref{fig:true_prior}a,][]{aminzadeh1997seg}, which is discretised using a regular 101 $\times$ 101 $\times$ 63 grid of cells with 50 m spacing. We deploy 81 sources (red dots in Figure \ref{fig:true_prior}a) and 10,201 receivers (yellow dots in Figure \ref{fig:true_prior}a) at the surface with regular spacings of 500 m and 50 m respectively. The waveform data are calculated using the time-domain finite difference method with a 2 to 10 Hz Ormsby wavelet \citep{ryan1994ricker}. Gradients of the likelihood function with respect to velocities are computed using the adjoint method \citep{tarantola1988theoretical, tromp2005seismic, fichtner2006adjoint, plessix2006review, liu2012seismic}.

\begin{figure}
	\includegraphics[width=1.\linewidth]{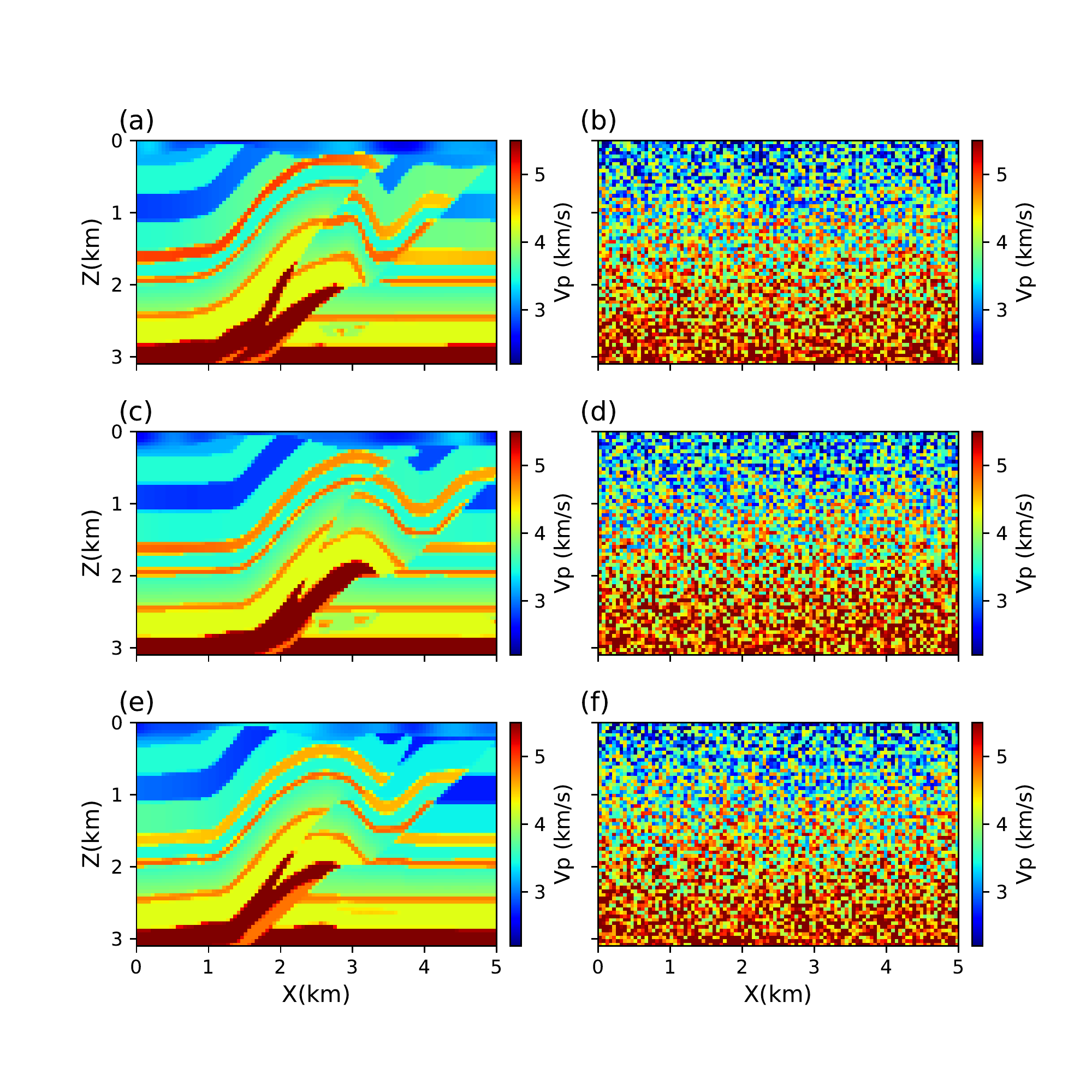}
	\caption{The true model (left column) and an initial particle (right column) at cross sections of Y = 1km (a and b), 2.5 km (c and d)  and 4 km (e and f), respectively.}
	\label{fig:true_init}
\end{figure}

We represent available prior information by a Uniform distribution over an interval width of 2.5 km/s at each depth (Figure \ref{fig:true_prior}b). Figure \ref{fig:true_init} shows a set of cross sections (Y=1km, 2.5km and 4km) of the true model and an example model generated from the prior distribution. For the likelihood function we assume that a Gaussian distribution with a diagonal covariance matrix can be used to represent uncertainties on the waveform data:
\begin{equation}
	p(\mathbf{d}_{\mathrm{obs}}|\mathbf{m}) \propto \mathrm{exp}\left[-\frac{1}{2}\sum_{i}\left(\frac{d_{i}^{\mathrm{obs}}-d_{i}(\mathbf{m})}{\sigma_{i}}\right)^{2}\right]
	\label{eq:likelihood}
\end{equation}
where $i$ denotes the index of time samples and $\sigma_{i}$ is the standard deviation of that data point. In this study we set $\sigma_{i}$ to be 2 percent of the median of the maximum amplitude of each seismic trace.

For ADVI we set the initial Gaussian distribution in the unconstrained space to be a standard Normal distribution $\mathcal{N}(\bm{\uptheta}|\mathbf{0},\mathbf{I})$, and update the distribution using the ADAM algorithm \citep{kingma2014adam} for 1,000 iterations after which point the average misfit across Monte Carlo samples ceases to decrease. To reduce the computational cost, we compute the gradients in equation (\ref{eq:gradient_mu}) and (\ref{eq:gradient_sigma}) using mini-batch data from 36 sources which are randomly selected from the total of 81 sources. At each iteration the gradients are calculated using four Monte Carlo samples. The final Gaussian distribution is transformed back to the original space, from which we generate 2,000 samples to visualize the results.

For SVGD we generate 400 particles from the prior distribution (an example is shown in Figure \ref{fig:true_init}), and transform them to an unconstrained space using equation (\ref{eq:transform}). Those particles are then updated using equation (\ref{eq:phi_star}) for 1,000 iterations after which point the average misfit across particles ceases to decrease. Similarly to above the gradients in equation (\ref{eq:phi_star}) are calculated using minibatch data from 36 sources. The final particles are transformed back to the original space.

For sSVGD we start from 20 particles that are generated from the prior distribution, and transform them to the unconstrained space as in SVGD. Those particles are then updated (sampled) using equation (\ref{eq:stochastic_svgd}) for 4,000 iterations with a burn-in period of 2,000. To reduce the memory and storage cost, we only retain every forth sample after the burn-in period. This results in a total of 10,000 samples, which are transformed back to the original space to calculate statistics of the estimated posterior pdf. At each iteration the gradients are also calculated using minibatch data from 36 sources. 

\subsection{Model comparison}
\begin{figure}
	\includegraphics[width=1.\linewidth]{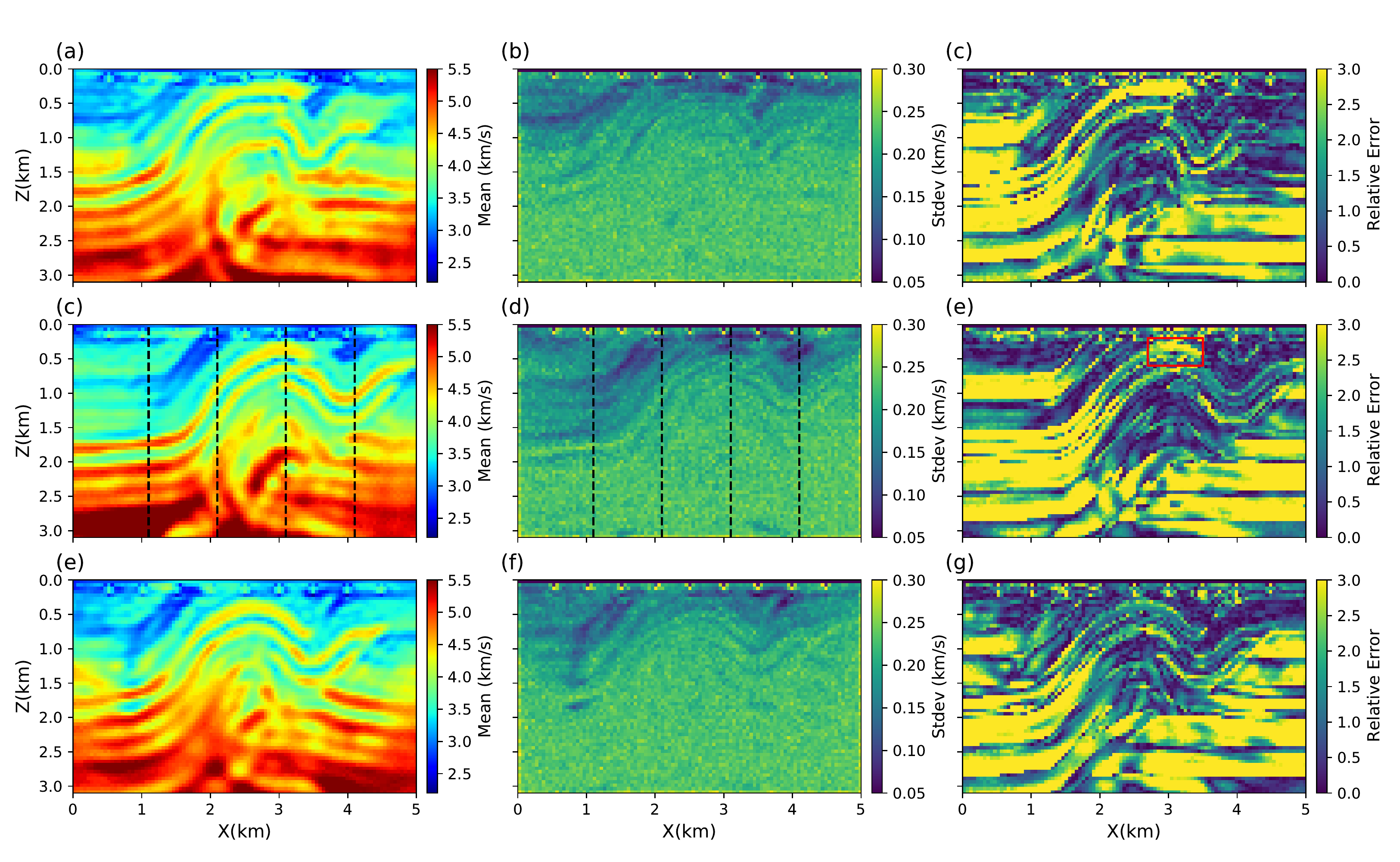}
	\caption{The mean (a, c and e), standard deviation (b, d and f) and relative error (c, e and g) obtained using ADVI over the same cross sections as in Figure \ref{fig:true_init}. The relative error is computed using $|\mathbf{m}^{mean}-\mathbf{m}^{true}|/\bm{\sigma}$ where $\bm{\sigma}$ is the standard deviation. Black dashed lines denote the well log locations referred to in the main text.}
	\label{fig:advi_results}
\end{figure}

Figure \ref{fig:advi_results} shows the mean, standard deviation and the relative error computed using $|\mathbf{m}^{mean}-\mathbf{m}^{true}|/\bm{\sigma}$ where $\bm{\sigma}$ is the standard deviation, obtained using ADVI, displayed on the same cross sections as in Figure \ref{fig:true_init}. In the shallow part (depth Z $<$ 1.5 km) the mean model shows similar structure to the true model. For example, overthrusted high velocity structures can be observed clearly in the mean model. Over the same depth range the standard deviation model shows similar features to the mean model. A similar phenomenon has been observed in a range of previous studies \citep{gebraad2020bayesian, zhang2020variational, zhang2021bayesiana}. At greater depths Z $>$ 1.5 km the mean model deviates from the true model. This is probably because of the lower sensitivity caused by the short source-receiver offset offered by our acquisition geometry. This is also supported by high uncertainties across the same area. The relative error shows that the deviation of the mean model from the true model is larger than three standard deviations at depth and on both sides, which suggests that the uncertainty is clearly underestimated there. This underestimation is likely caused by the mean-field approximation we have used in ADVI (see Figure \ref{fig:mean-field}a).

\begin{figure}
	\includegraphics[width=1.\linewidth]{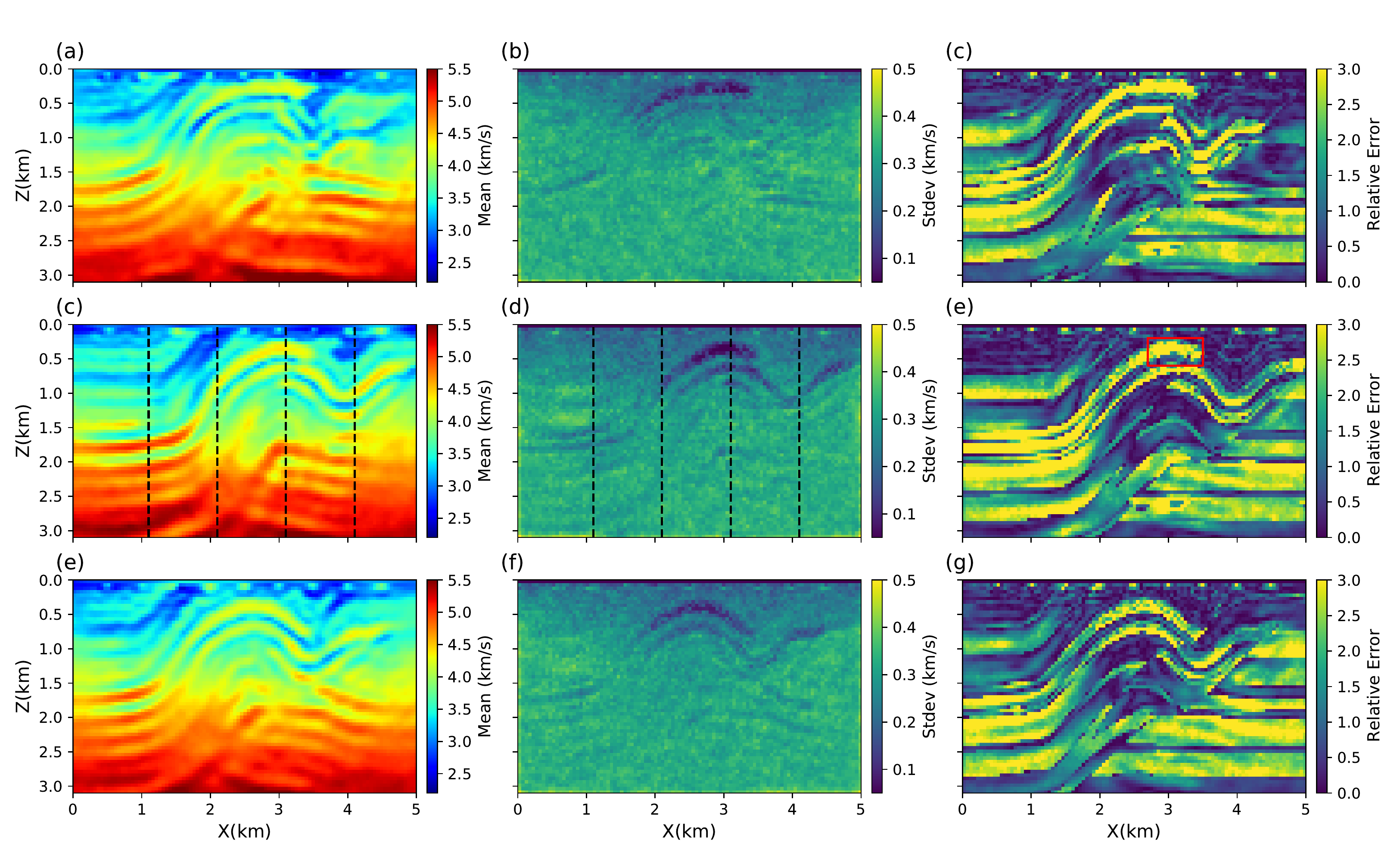}
	\caption{The mean, standard deviation and relative error obtained using SVGD. Key as in Figure \ref{fig:advi_results}.}
	\label{fig:svgd_results}
\end{figure}

\begin{figure}
	\includegraphics[width=1.\linewidth]{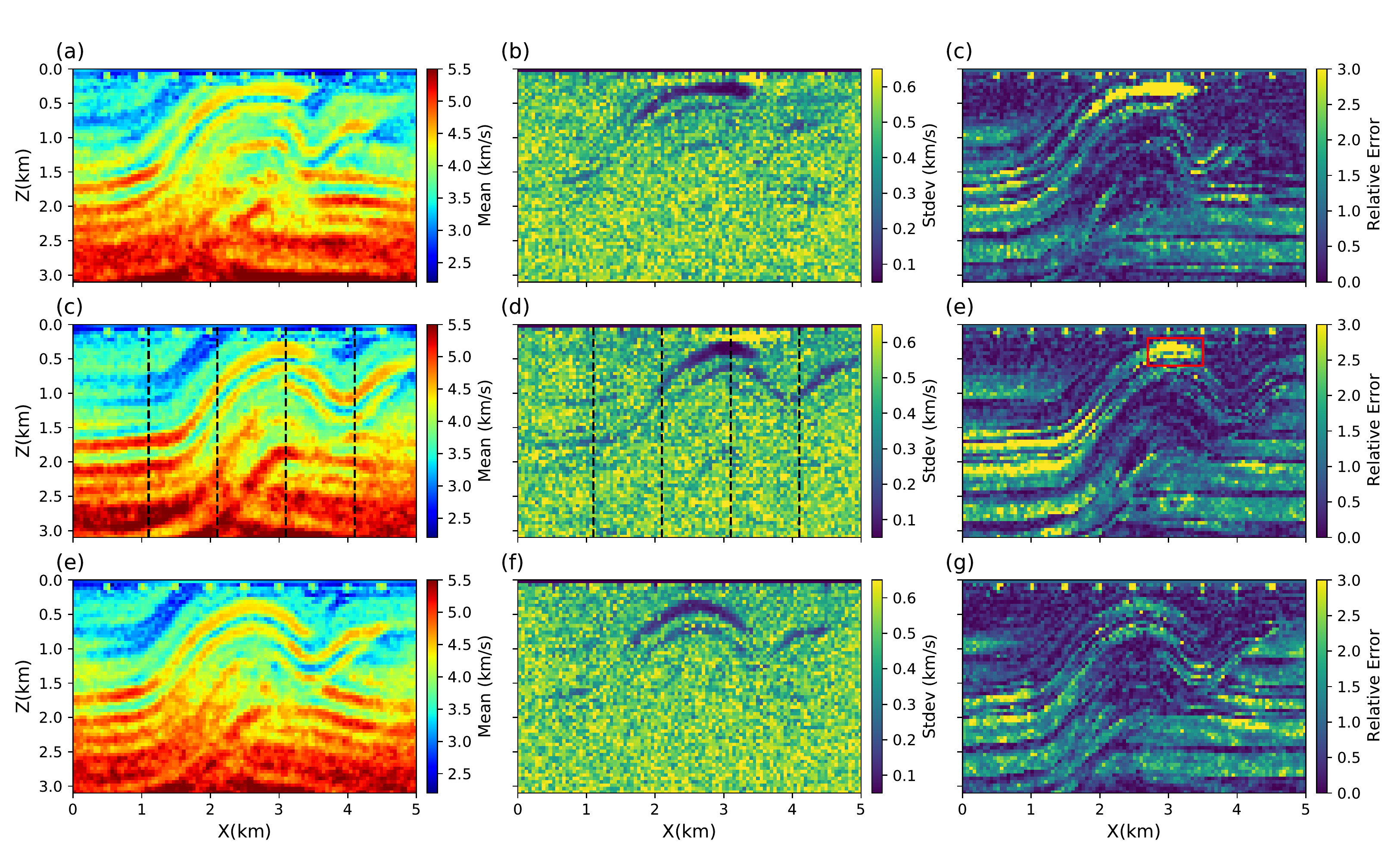}
	\caption{The mean, standard deviation and relative error obtained using sSVGD. Key as in Figure \ref{fig:advi_results}.}
	\label{fig:ssvgd_results}
\end{figure}

\begin{figure}
	\includegraphics[width=1.\linewidth]{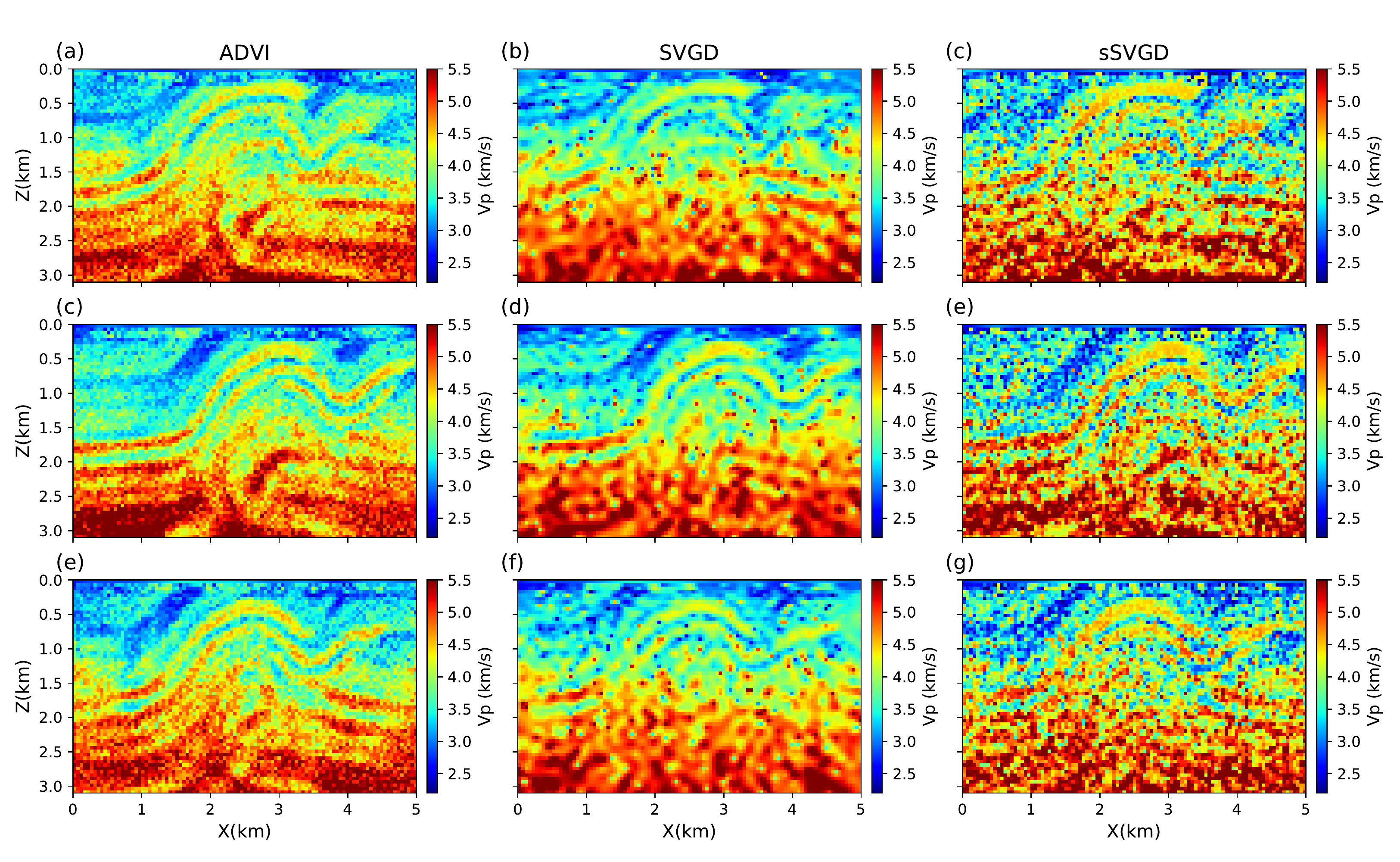}
	\caption{Example particles obtained using ADVI (a, c and e), SVGD (b,d and f) and sSVGD (c, e and g) over the same cross sections as in Figure \ref{fig:true_init}.}
	\label{fig:sample_results}
\end{figure}

Figure \ref{fig:svgd_results} shows the results obtained using SVGD. Overall the results show similar mean and standard deviation structures to those obtained using ADVI. For example, the mean model shows similar features to the true model in the shallow parts, and deviates from the true model at greater depths. The standard deviation also shows similar features to the mean model across the shallow part and higher uncertainties at greater depths. Note that the magnitude of the standard deviations is generally higher than those obtained using ADVI, which again shows the limitation of the mean-field approximation. However, although the relative error is smaller than those from ADVI, there is still a large part of the model whose relative error is higher than three standard deviations which suggests that SVGD can also underestimate the uncertainty \citep{ba2021understanding}. This is probably because we use a small number of particles (400) to represent a probability distribution in an extremely high dimensional space (642,663). Consequently for those parts that are not well constrained by the data which should have a broader posterior distribution, it becomes impossible to represent the posterior distribution. Although the results can be further improved by using a larger number of particles \citep{zhang2021introduction}, this incurs a significantly higher computational cost.

Figure \ref{fig:ssvgd_results} shows the results obtained using sSVGD. Ignoring magnitudes for the moment, the overall shapes of the mean and standard deviation models are similar to those obtained using ADVI and SVGD suggesting that these shapes may be reliable for this specific problem. Note that the mean model obtained using sSVGD is more similar to the true model, which may indicate that sSVGD produced more accurate results than ADVI or SVGD as we have discussed in section 2. In addition, the magnitudes of the standard deviation are much higher than those obtained using ADVI or SVGD, and the relative error is also significantly smaller. For most parts the relative error obtained using sSVGD is smaller than three standard deviations, which is again indicative of the higher accuracy of sSVGD compared to ADVI or SVGD. Similarly to previous results, the deeper parts and two sides show larger errors than the rest of the model because of the lower sensitivity  of our data to those parts. Note that the results obtained using ADVI and SVGD show smoother structures than those obtained using sSVGD. This is because in ADVI and SVGD the results are obtained deterministically, whereas sSVGD is a stochastic McMC method which therefore represents more randomness. A similar phenomenon was observed by \cite{zhang2020variational} when comparing results obtained using SVGD and HMC. We also note that the results can be further improve by running the sSVGD for longer.

In Figure \ref{fig:sample_results} we show examples of samples (particles) obtained using each method at the same cross sections as above. Overall the samples obtained using different methods show similar structures. For example, the shallow part (Z $<$ 1.5 km) shows similar features to the true model, whereas the deeper part has more random structures. Similarly to the mean and standard deviation models, the sample obtained using SVGD is smoother than that obtained using sSVGD. There is no correlation between parameters in ADVI, so the sample obtained using ADVI shows random structures at pixel scale. 

\begin{figure}
	\includegraphics[width=.9\linewidth]{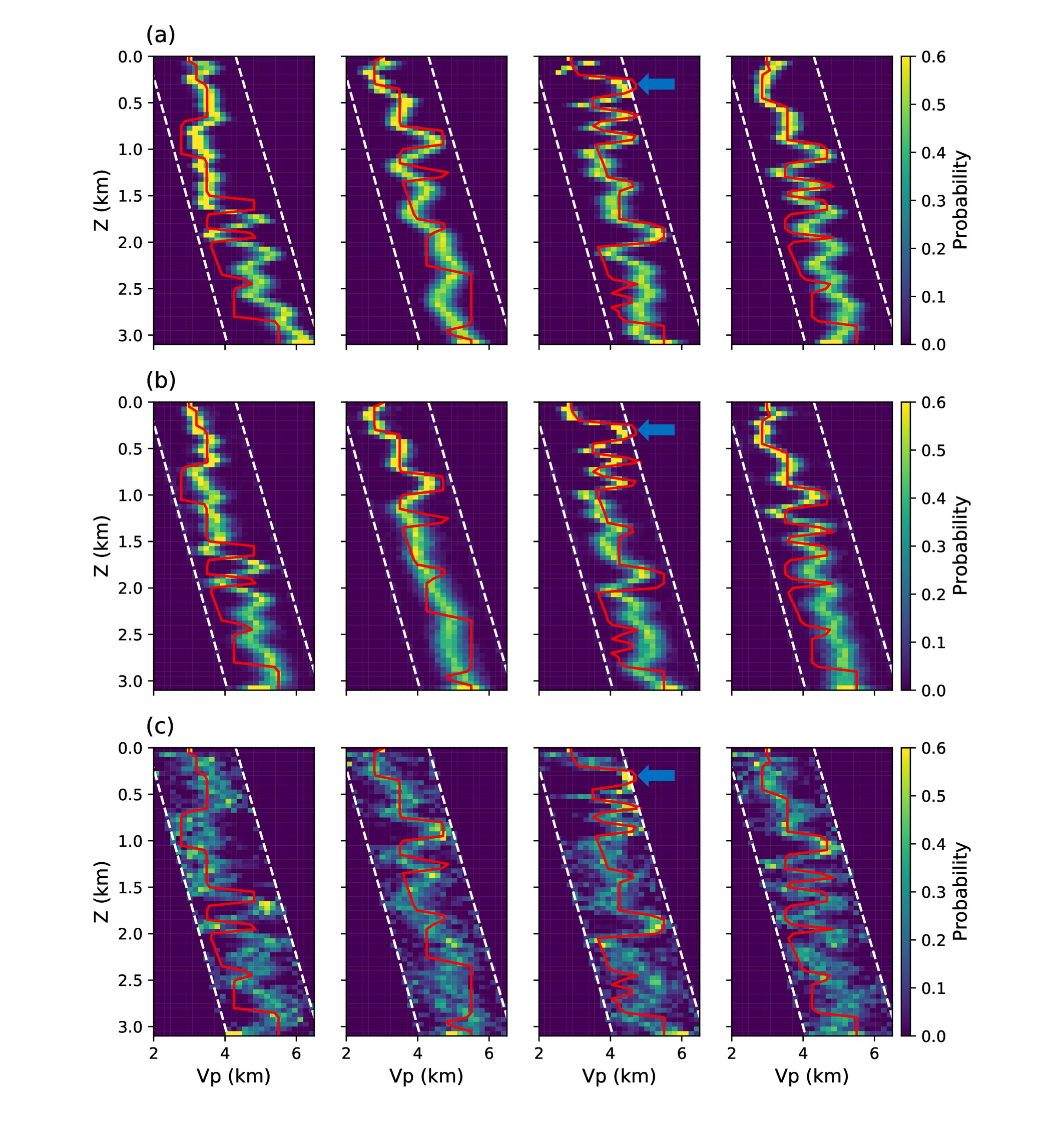}
	\caption{The marginal distributions at four well logs (black dashed line in Figure \ref{fig:advi_results}, \ref{fig:svgd_results} and \ref{fig:ssvgd_results}) obtained using (a) ADVI, (b) SVGD and (c) sSVGD respectively. Red lines show the true velocity profiles and white dashed lines show the lower and upper bound of the prior distribution.}
	\label{fig:marginals}
\end{figure}

To further analyse the results, we show the marginal distributions obtained using the suite of methods along four vertical profiles simulating well logs, whose locations are indicated using black dashed lines in Figures \ref{fig:advi_results}, \ref{fig:svgd_results} and \ref{fig:ssvgd_results}. The results clearly show that the marginal distributions obtained using sSVGD are wider than those obtained using ADVI and SVGD as we have already observed. Across deeper parts (Z $>$ 1.5 km), the true velocity values lie outside of the high probability area in the results obtained using ADVI and SVGD (Figure \ref{fig:marginals}a and b), which again demonstrates that ADVI and SVGD can underestimate uncertainty. In contrast, sSVGD produces more reasonable uncertainty estimates since they at least generally include the true model in values with non-zero uncertainty. Overall the results show lower uncertainty in the shallower part (Z $<$ 1.5 km) and higher uncertainty at the deeper part as we expect. Note that at the depth of 0.4 km in the third well log (denoted by a blue arrow), the marginal distributions concentrate close to the upper bound of the prior distribution. This is because the true velocity at this location is higher than the prior upper bound, which also explains the large relative error in this area (red box in Figure \ref{fig:advi_results}, \ref{fig:svgd_results} and \ref{fig:ssvgd_results}). This result provides useful insight into the performance of these methods in real applications as it is not uncommon to impose inappropriate prior information in practice.

\subsection{Computational cost}  

In Table \ref{tb:cost} we summarize the number of simulations, the number of CPU cores, and the wall clock time required by each method. The number of simulations provides a good metric of the overall computational cost as for each method the forward and adjoint simulations are the most time-consuming components of these calculations. Given that all of the methods can be fully parallelised, for example, the gradient calculation in each method can be performed independently for each particle (sample), the number of CPU cores together with the wall clock time provide additional insights into the computational requirement in practice. 

 \begin{table}
 \begin{center}
 \begin{minipage}{120mm}
	\caption{A comparison of computational cost for the 3 inference methods.} 	
	\begin{tabular}{l c c c}
		\hline
		Method  & Number of simulations\textsuperscript{a} & CPU cores \textsuperscript{b} & Wall time (hours) \\
		\hline
		ADVI  & 4,000 & 768 & 53.8 \\
		SVGD & 400,000 & 7680 & 558.7 \\
		sSVGD & 80,000 & 3840 & 220.8 \\
		\hline
		\multicolumn{4}{l}{$^{a}$This is measured as the number of minibatch simulations.} \\
		\multicolumn{4}{l}{$^{b}$The CPU used in this study is Intel Xeon Platinum.}
	\end{tabular}
	\label{tb:cost}
\end{minipage}
\end{center}
\end{table}

The results show that ADVI is the cheapest method as it only requires 4,000 simulations which we performed using 768 CPU cores, but we have demonstrated above that the method is likely to produce systematically biased results. However, given that the method is extremely efficient (only requiring 53.8 hours in real time), ADVI could still be used to provide a first, relatively rapid insight into the subsurface structure. In addition, as we have demonstrated in Figure \ref{fig:mean-field}a, the method can be used to provide a lower bound estimate of the uncertainty. SVGD appears to be the most expensive method, which requires 400,000 simulations and takes approximately 23 days to run using 7,680 CPU cores. Because of the limited number of particles the method also provides biased results as we have shown above, which makes SVGD a less attractive method for 3D FWI in practice. In contrast, by adding a noise term to the dynamics of SVGD, sSVGD can use a small number of particles to generate many final model samples, which makes the method relatively efficient. For example, to obtain the above results sSVGD required five times fewer simulations than SVGD. However, because of the randomness introduced by the noise term, sSVGD requires more iterations to converge which makes the method only 2$\sim$3 times more efficient in real time. Given that sSVGD also provides the most accurate results among the three methods, the method would be a good choice for practical applications. In addition, since it is a McMC method the results of sSVGD can always be improved by performing more iterations, whereas the same method of improvement cannot be employed when using ADVI or SVGD.

Note that the above comparison depends on subjective assessments of the point of convergence for each method, so the absolute computational time may not be entirely accurate. Nevertheless the comparison at least provides a reasonable insight into the efficiency of each method. We also note that all of the methods require computation of gradients, which in this study are calculated efficiently using adjoint methods. For situations in which gradients are expensive to compute, the above suite of methods may become less efficient, and in such cases other methods that do not require gradients may be preferred.

 
\section{Discussion}
The primary result of this work is to show that variational methods (ADVI, SVGD and sSVGD) can be used to solve 3D Bayesian FWI problems. For ADVI, we used a mean-field approximation to reduce the computational cost, which systematically underestimates the uncertainty. To further improve the results, a full-rank covariance matrix may be used if sufficient computational resources are available, or a sparse covariance matrix which only includes correlation information between neighbouring cells can be implemented. ADVI minimizes KL$[q||p]$ to estimate the posterior distribution which can provide a lower bound estimate of the uncertainty in the mean-field case. On the other hand, methods such as the expectation propagation \citep{minka2013expectation} which minimizes KL$[p||q]$ instead of KL$[q||p]$, may be used to provide an upper bound estimate of the uncertainty.

We have demonstrated that for 3D FWI SVGD can provide biased results because of the limited number of particles. Instead of increasing the number of particles which may be computationally intractable, one may try to reduce the dimensionality of the problem. For example, other parameterizations that require fewer parameters to represent the model may be used, such as Voronoi cells \citep{bodin2009seismic, zhang20183}, wavelet parameterization \citep{hawkins2015geophysical}, Johnson-Mehl tessellation \citep{belhadj2018new}, Delaunay and Clough-Tocher parameterizations \citep{curtis1997reconditioning} or discrete cosine transforms \citep{urozayev2022reduced}. In addition, other SVGD variants which project the high dimensional parameter space into a lower dimensional space may be used to improve the results, for example, projected SVGD \citep{chen2020projected} or sliced SVGD \citep{gong2020sliced}.

By adding a noise term to the dynamics of SVGD, sSVGD becomes a McMC method with multiple interactive chains. Note that this is different from other McMC methods which run multiple interactive chains such as parallel tempering \citep{earl2005parallel, sambridge2013parallel}. In parallel tempering, a set of chains with different temperatures are run in parallel, and at each iteration samples in two randomly selected (or neighbouring) chains are exchanged with a Metropolis-Hastings criterion. In sSVGD, all Markov chains interact by using a kernel function and hence no sample exchange occurs between chains. 

Although sSVGD provides more accurate results than ADVI and SVGD, it also requires more iterations to converge. To improve efficiency of the method, one might exploit higher order gradient information, for example, using a Hessian matrix kernel \citep{wang2019stein} or the stochastic Stein variational Newton method \citep{leviyev2022stochastic}. Since sSVGD is a McMC method, one can further improve the accuracy of the method by implementing a Metropolis-Hastings correction step at each iteration \citep{metropolis1949monte, hastings1970monte}, though in such cases stochastic minibatches may not be used because of the detailed balance requirement of the Metropolis-Hastings step. 

Note that for both SVGD and sSVGD, the posterior distribution is likely to be under sampled given the large dimensionality (642,663) and the small number of samples (400 and 10,000 respectively). While the set of samples may not be sufficient to represent the full posterior distribution, they may at least provide reasonable mean and (in the case of sSVGD) standard deviation estimates. We also note that in practice the number of samples is always restricted by the available computational cost.

In this study we used a Uniform prior distribution. This may cause posterior pdfs to occur that are more complex than would be the case if Gaussian or other prior distributions were used that more strongly focus the solution towards certain regions of parameter space. This means that our posterior pdf may be harder to explore than would otherwise be the case. In practice where more knowledge about the subsurface is available, one can use a more informative prior distribution. For example, models obtained using fast travel time tomography can be used as prior information for FWI. In addition, prior regularization or Gaussian processes may be used to produce smoother models \citep{mackay2003information, ray2019bayesian}. Neural networks can also be used to encode geological information into prior distributions \citep{laloy2017inversion, mosser2020stochastic}.

For the likelihood function we used Gaussian data uncertainties with a known, fixed data noise level. In practice this noise level should be determined from the data, for example, by using the maximum likelihood method \citep{sambridge2013parallel}. It may also be possible to estimate the noise level in the inversion process using a hierarchical Bayesian formulation \citep{ranganath2016hierarchical, malinverno2004expanded}. We also note that other non-Gaussian likelihood functions may be used to improve the results given that those likelihood functions are defined to represent the probability distribution of data uncertainty \citep{zhang2022surface}.

For computational efficiency we only applied the methods to a small area with a small dataset. In practice, the methods may become computationally intractable for large subsurface volumes and large datasets due to the curse of dimensionality \citep{curtis2001prior}. In such cases one may use experimental design methods \citep{curtis2004seismic, maurer2010recent} to select a small part of the large dataset, and perform inversions using those selected data. Faster, approximate forward modelling methods may also be used to improve efficiency of the methods, for example neural network based modelling methods \citep{sirignano2018dgm}. We also note that apart from the mean and uncertainty models, the obtained samples can be used for real-world applications, for example, providing models for reservoir simulations or answering specific scientific questions \citep{arnold2018interrogation, zhang2022interrogating, zhao2022interrogating}.
 
\section{Conclusion}
In this study we applied three different variational inference methods: automatic differential variational inference (ADVI), Stein variational gradient descent (SVGD) and stochastic SVGD (sSVGD) to 3D full-waveform inversion, and demonstrated feasibility of using these methods to solve large scale probabilistic inverse problems. The results show that ADVI with a mean-field approximation can provide rapid solutions but with systematically underestimated uncertainty. In practice, the method can therefore be used to provide a rapid initial estimate of the solution, or to provide a lower bound estimate of the uncertainty. SVGD appears to be the most expensive method, but still provides a biased solution because of the limited number of particles. In contrast, by adding a noise term in the dynamics of SVGD, sSVGD becomes a Markov chain Monte Carlo method and provides the most accurate results. We thus conclude that variational inference methods can be used to solve real-world 3D full wave form inversion problems.   
 
\begin{acknowledgments}
The authors thank the Edinburgh Imaging Project sponsors (BP and Total) for supporting this research.
\end{acknowledgments}

\bibliographystyle{gji}
\bibliography{../bibliography}

\appendix

\label{lastpage}

\end{document}